
\noindent
%
\newbox\hdbox%
\newcount\hdrows%
\newcount\multispancount%
\newcount\ncase%
\newcount\ncols
\newcount\nrows%
\newcount\nspan%
\newcount\ntemp%
\newdimen\hdsize%
\newdimen\newhdsize%
\newdimen\parasize%
\newdimen\thicksize%
\newdimen\thinsize%
\newdimen\tablewidth%
\newif\ifcentertables%
\newif\ifendsize%
\newif\iffirstrow%
\newif\iftableinfo%
\newtoks\dbt%
\newtoks\hdtks%
\newtoks\savetks%
\newtoks\tableLETtokens%
\newtoks\tabletokens%
\newtoks\widthspec%
%
%
\immediate\write15{%
-----> TABLE MACROS LOADED%
}%
%
%
\tableinfotrue%
\catcode`\@=11
\def\tstrut{\vrule height16pt depth6pt width0pt}%
\def\|{|}
\def\tablerule{\noalign{\hrule height\thinsize depth0pt}}%
\thicksize=1.5pt
\thinsize=0.6pt
\def\thickrule{\noalign{\hrule height\thicksize	depth0pt}}%
\def\ctr#1{\hfil\ #1\hfil}%
%
%
%
\tablewidth=-\maxdimen%
\def\tabskipglue{0pt plus 1fil minus 1fil}%
%
%
\centertablestrue%
%
%
%
%
\parasize=4in%
\gdef\ARGS{########}
\gdef\headerARGS{####}
\def\@mpersand{&}
{\catcode`\|=13
\gdef\letbarzero{\let|0}
\gdef\letbartab{\def|{&&}}
}
{\def\ampskip{&\omit\hfil&}
\catcode`\&=13
\let&0
\xdef\letampskip{\def&{\ampskip}}
}
\def\begintable{
   \begingroup%
   \catcode`\|=13\letbartab%
   \catcode`\&=13\letampskip%
   \def\multispan##1{
      \omit \mscount##1%
      \multiply\mscount\tw@\advance\mscount\m@ne%
      \loop\ifnum\mscount>\@ne \sp@n\repeat%
   }
   \def\|{%
      &\omit\widevline&%
   }%
   \ruledtable
}
\long\def\ruledtable#1\endtable{%
%
%
%
   \offinterlineskip
   \tabskip 0pt
   \def\widevline{\vrule width\thicksize}
   \def\endrow{\@mpersand\omit\hfil\crnorm\@mpersand}%
   \def\crthick{\@mpersand\crnorm\thickrule\@mpersand}%
   \def\crnorule{\@mpersand\crnorm\@mpersand}%
   \let\nr=\crnorule
   \def\endtable{\@mpersand\crnorm\thickrule}%
   \let\crnorm=\cr
%
%
   \edef\cr{\@mpersand\crnorm\tablerule\@mpersand}%
   \the\tableLETtokens
%
%
   \tabletokens={&#1}
%
%
   \countROWS\tabletokens\into\nrows%
   \countCOLS\tabletokens\into\ncols%
%
%
   \advance\ncols by -1%
   \divide\ncols by 2%
   \advance\nrows by 1%
%
%
   \iftableinfo	%
      \immediate\write16{[Nrows=\the\nrows, Ncols=\the\ncols]}%
   \fi%
%
%
   \ifcentertables
      \line{
      \hss
   \else %
      \hbox{%
   \fi
      \vbox{%
	 \makePREAMBLE{\the\ncols}
	 \edef\next{\preamble}
	 \let\preamble=\next
	 \makeTABLE{\preamble}{\tabletokens}
      }
      \ifcentertables \hss}\else }\fi
   \endgroup
   \tablewidth=-\maxdimen
}
\def\makeTABLE#1#2{
   {
   \let\ifmath0
   \let\header0
   \let\multispan0
%
%
   \ifdim\tablewidth>-\maxdimen	%
 \widthspec=\expandafter{\expandafter t\expandafter o%
 \the\tablewidth}%
   \else %
      \widthspec={}%
   \fi %
   \xdef\next{
      \halign\the\widthspec{%
      #1
      \noalign{\hrule height\thicksize depth0pt}
      \the#2\endtable
%
      }
   }
   }
   \next
}
\def\makePREAMBLE#1{
   \ncols=#1
   \begingroup
   \let\ARGS=0
   \edef\xtp{\widevline\ARGS\tabskip\tabskipglue%
   &\tstrut\ctr{\ARGS}}
   \advance\ncols by -1
   \loop
      \ifnum\ncols>0 %
      \advance\ncols by	-1%
      \edef\xtp{\xtp&\vrule width\thinsize\ARGS&\ctr{\ARGS}}%
   \repeat
   \xdef\preamble{\xtp&\widevline\ARGS\tabskip0pt%
   \crnorm}
   \endgroup
}
\def\countROWS#1\into#2{
   \let\countREGISTER=#2%
   \countREGISTER=0%
   \expandafter\ROWcount\the#1\endcount%
}%
\def\ROWcount{%
   \afterassignment\subROWcount\let\next= %
}%
\def\subROWcount{%
   \ifx\next\endcount %
      \let\next=\relax%
   \else%
      \ncase=0%
      \ifx\next\cr %
	 \global\advance\countREGISTER by 1%
	 \ncase=0%
      \fi%
      \ifx\next\endrow %
	 \global\advance\countREGISTER by 1%
	 \ncase=0%
      \fi%
      \ifx\next\crthick	%
	 \global\advance\countREGISTER by 1%
	 \ncase=0%
      \fi%
      \ifx\next\crnorule %
	 \global\advance\countREGISTER by 1%
	 \ncase=0%
      \fi%
      \ifx\next\header %
	 \ncase=1%
      \fi%
      \relax%
      \ifcase\ncase %
	 \let\next\ROWcount%
      \or %
	 \let\next\argROWskip%
      \else %
      \fi%
   \fi%
   \next%
}
\def\counthdROWS#1\into#2{%
\dvr{10}%
   \let\countREGISTER=#2%
   \countREGISTER=0%
\dvr{11}%
\dvr{13}%
   \expandafter\hdROWcount\the#1\endcount%
\dvr{12}%
}%
\def\hdROWcount{%
   \afterassignment\subhdROWcount\let\next= %
}%
\def\subhdROWcount{%
   \ifx\next\endcount %
      \let\next=\relax%
   \else%
      \ncase=0%
      \ifx\next\cr %
	 \global\advance\countREGISTER by 1%
	 \ncase=0%
      \fi%
      \ifx\next\endrow %
	 \global\advance\countREGISTER by 1%
	 \ncase=0%
      \fi%
      \ifx\next\crthick	%
	 \global\advance\countREGISTER by 1%
	 \ncase=0%
      \fi%
      \ifx\next\crnorule %
	 \global\advance\countREGISTER by 1%
	 \ncase=0%
      \fi%
      \ifx\next\header %
	 \ncase=1%
      \fi%
\relax%
      \ifcase\ncase %
	 \let\next\hdROWcount%
      \or%
	 \let\next\arghdROWskip%
      \else %
      \fi%
   \fi%
   \next%
}%
{\catcode`\|=13\letbartab
\gdef\countCOLS#1\into#2{%
   \let\countREGISTER=#2%
   \global\countREGISTER=0%
   \global\multispancount=0%
   \global\firstrowtrue
   \expandafter\COLcount\the#1\endcount%
   \global\advance\countREGISTER by 3%
   \global\advance\countREGISTER by -\multispancount
}%
\gdef\COLcount{%
   \afterassignment\subCOLcount\let\next= %
}%
{\catcode`\&=13%
\gdef\subCOLcount{%
   \ifx\next\endcount %
      \let\next=\relax%
   \else%
      \ncase=0%
      \iffirstrow
	 \ifx\next& %
	    \global\advance\countREGISTER by 2%
	    \ncase=0%
	 \fi%
	 \ifx\next\span	%
	    \global\advance\countREGISTER by 1%
	    \ncase=0%
	 \fi%
	 \ifx\next| %
	    \global\advance\countREGISTER by 2%
	    \ncase=0%
	 \fi
	 \ifx\next\|
	    \global\advance\countREGISTER by 2%
	    \ncase=0%
	 \fi
	 \ifx\next\multispan
	    \ncase=1%
	    \global\advance\multispancount by 1%
	 \fi
	 \ifx\next\header
	    \ncase=2%
	 \fi
	 \ifx\next\cr	    \global\firstrowfalse \fi
	 \ifx\next\endrow   \global\firstrowfalse \fi
	 \ifx\next\crthick  \global\firstrowfalse \fi
	 \ifx\next\crnorule \global\firstrowfalse \fi
      \fi
\relax
      \ifcase\ncase %
	 \let\next\COLcount%
      \or %
	 \let\next\spancount%
      \or %
	 \let\next\argCOLskip%
      \else %
      \fi %
   \fi%
   \next%
}%
\gdef\argROWskip#1{%
   \let\next\ROWcount \next%
}
\gdef\arghdROWskip#1{%
   \let\next\ROWcount \next%
}
\gdef\argCOLskip#1{%
   \let\next\COLcount \next%
}
}
}
\def\spancount#1{
   \nspan=#1\multiply\nspan by 2\advance\nspan by -1%
   \global\advance \countREGISTER by \nspan
   \let\next\COLcount \next}%
\def\dvr#1{\relax}%
\def\header#1{%
\dvr{1}{\let\cr=\@mpersand%
\hdtks={#1}%
\counthdROWS\hdtks\into\hdrows%
\advance\hdrows	by 1%
\ifnum\hdrows=0	\hdrows=1 \fi%
\dvr{5}\makehdPREAMBLE{\the\hdrows}%
\dvr{6}\getHDdimen{#1}%
{\parindent=0pt\hsize=\hdsize{\let\ifmath0%
\xdef\next{\valign{\headerpreamble #1\crnorm}}}\dvr{7}\next\dvr{8}%
}%
}\dvr{2}}
\def\makehdPREAMBLE#1{
\dvr{3}%
\hdrows=#1
{
\let\headerARGS=0%
\let\cr=\crnorm%
\edef\xtp{\vfil\hfil\hbox{\headerARGS}\hfil\vfil}%
\advance\hdrows	by -1
\loop
\ifnum\hdrows>0%
\advance\hdrows	by -1%
\edef\xtp{\xtp&\vfil\hfil\hbox{\headerARGS}\hfil\vfil}%
\repeat%
\xdef\headerpreamble{\xtp\crcr}%
}
\dvr{4}}
\def\getHDdimen#1{%
\hdsize=0pt%
\getsize#1\cr\end\cr%
}
\def\getsize#1\cr{%
\endsizefalse\savetks={#1}%
\expandafter\lookend\the\savetks\cr%
\relax \ifendsize \let\next\relax \else%
\setbox\hdbox=\hbox{#1}\newhdsize=1.0\wd\hdbox%
\ifdim\newhdsize>\hdsize \hdsize=\newhdsize \fi%
\let\next\getsize \fi%
\next%
}%
\def\lookend{\afterassignment\sublookend\let\looknext= }%
\def\sublookend{\relax%
\ifx\looknext\cr %
\let\looknext\relax \else %
   \relax
   \ifx\looknext\end \global\endsizetrue \fi%
   \let\looknext=\lookend%
    \fi	\looknext%
}%
%
%
\def\tablelet#1{%
   \tableLETtokens=\expandafter{\the\tableLETtokens #1}%
}%
\catcode`\@=12

\def\etal{{\it et al.}~~}
\def\cf{{\it cf.}~}
\font\machm=cmsy10 \skewchar\machm='60

\def\refset{\parindent=0pt\hangafter=1\hangindent=1em}
\magnification=1200
\hsize=5.80truein
\hoffset=1.20truecm
\newcount\eqtno
\eqtno = 1
\parskip 3pt plus 1pt minus .5pt
\baselineskip 21.5pt plus .1pt
%
%
%
%
\centerline{ \  }
\vskip 0.35in
\centerline{\bf X-RAY CLUSTERS IN A CDM$+\Lambda$ UNIVERSE:}
\centerline{\bf A DIRECT, LARGE-SCALE, HIGH RESOLUTION, HYDRODYNAMIC
SIMULATION}
\vskip 1.5cm
\centerline{Renyue Cen and Jeremiah P. Ostriker}
\centerline{\it Princeton University Observatory}
\centerline{\it Princeton, NJ 08544 USA}
\vskip 0.7cm
\centerline{Submitted to {\it The Astrophysical Journal} on Sept 30, 1993}
\centerline{Dec 22, 1993}
\vfill\eject

\centerline{\bf ABSTRACT}
A new, three-dimensional, shock capturing, hydrodynamic code
is utilized to determine the distribution of hot gas in
a CDM$+\Lambda$ model universe.
Periodic boundary conditions are assumed:
a box with size $85h^{-1}$Mpc,
having cell size
$0.31h^{-1}$Mpc, is followed in a simulation with $270^3=10^{7.3}$ cells.
We adopt $\Omega=0.45$, $\lambda=0.55$, $h\equiv H/100$km/s/Mpc$=0.6$,
and then, from COBE and light element nucleosynthesis,
$\sigma_8=0.77$, $\Omega_b=0.043$.
We identify the X-ray emitting clusters in the simulation box,
compute the luminosity function at several wavelength bands,
the temperature function and estimated sizes,
as well as the evolution of these quantities with redshift.
This open model succeeds in matching local observations
of clusters in contrast to the standard $\Omega=1$, CDM model,
which fails.
It predicts an order of magnitude decline
in the number density of bright ($h\nu = 2-10$keV)
clusters from $z=0$ to $z=2$
in contrast to
a slight increase in the number density
for standard $\Omega=1$, CDM model.

This COBE-normalized CDM$+\Lambda$ model produces
approximately the same number of X-ray clusters having
$L_x>10^{43}$erg/s as observed.
The background radiation field
at 1keV due to clusters
is approximately $10\%$ of the observed background which,
after correction for numerical effects,
again indicates that the model is consistent with observations.

The number density of bright clusters
increases to $z\sim 0.2-0.5$
and declines,
but the luminosity per typical cluster decreases
monotonically with redshift,
with the result that the number density
of bright clusters
shows a broad peak near $z=0.5$,
and then a rapid decline as $z\rightarrow 3$.
The most interesting point which we find
is that
the temperatures of clusters in this model
freeze out at later times ($z\le 0.3$),
while previously we found in the CDM model
that there was
a steep increase during the same interval of redshift.
Equivalently, we find that $L^*$ of the Schechter fits of
cluster luminosity functions peaks near $z=0.3$ in this model,
while in the CDM model it is a monotonically decreasing function
of redshift.
Both trends
should be detectable even with a relatively ``soft" X-ray instrument
such as ROSAT, providing
a powerful discriminant between $\Omega=1$ and $\Omega<1$ models.
Detailed computations of the luminosity functions in the range
$L_x=10^{40}-10^{44}$erg/s in various energy
bands are presented for both cluster cores ($r\le 0.5h^{-1}$Mpc)
and total luminosities ($r<1h^{-1}$Mpc).
These are to be
used for comparison with ROSAT and other observational data sets.
They show the above noted negative evolution for
the open model.

We find little dependence of core radius on cluster luminosity
and the dependence of temperature on luminosity
$\log kT_x = A+B\log L_x$,
which is slightly steeper ($B=0.32\pm 0.01$)
than indicated
by observations ($B=0.265\pm 0.035$),
but within
observational errors.
In contrast, the standard $\Omega=1$ model
predicted temperatures which were significantly too high.
The mean luminosity-weighted temperature
is $1.8$keV, dramatically lower (by a factor of $3.5$)
than that found in the $\Omega=1$ model, and the evolution far
slower ($-30\%$ vs $-50\%$) than
in the $\Omega=1$ model to redshift $z=0.5$.
A modest average temperature gradient is found with temperatures
dropping to $90\%$ of central values
at $0.4h^{-1}$Mpc and to $60\%$ of central values
at $0.9h^{-1}$Mpc.

Examining the ratio of gas-to-total mass
in the clusters,
we find a slight antibias
($b=0.9$ or $({\Omega_{gas}\over\Omega_{tot}})_{cl}=0.083\pm 0.007$),
which is consistent
with observations
[$({\Omega_{gas}\over\Omega_{tot}})_{obs}=0.097\pm 0.019$
for the Coma cluster for the given value of $h$, \cf, White 1991].

\noindent
Cosmology: large-scale structure of Universe
-- hydrodynamics
-- Radiation Mechanisms: Bremsstrahlung
-- X-ray: general
\vfill\eject

\centerline{1. INTRODUCTION}

In the preceding two papers of this series
(Kang \etal 1994 = KCOR hereafter; Bryan \etal 1994),
we have examined the properties of X-ray clusters in
the standard COBE-normalized CDM universe and
reached two primary conclusions:
1) the standard
CDM model overproduces bright X-ray clusters ($L_x>10^{43}$erg/s)
by a factor in excess of five;
2) observations of the ratio
of gas-to-total mass in great clusters of galaxies,
combined with our simulations,
impliy that we live in an open ($\Omega=0.2$ to $0.3$)
universe (with or without a cosmological constant).
There are other well-known difficulties
with the standard CDM scenario (\cf Ostriker 1993 for a review),
and there are
advantages to open CDM models (\cf Efstathiou 1992)
which arise from other considerations,
independent of cluster X-ray properties.
Motivated by this knowledge,
we turn in this paper to examine
an open CDM model with a cosmological constant.
We wish to investigate the difference
in the evolutionary behavior as well as the $z=0$ properties
of X-ray clusters
in these two different models.
There is every expectation that
cluster X-ray properties should provide a strong
discriminant among cosmic theories.
We will, throughout this paper, make comparisons
with KCOR wherever it is possible, since that calculation was made
with identical physical assumptions
and with identical numerical modeling
techniques adopted.

In \S 2 we outline the method and initial conditions,
in \S 3 the results, and in \S 4 we assemble our conclusions.

\medskip
\centerline{2. METHOD AND INITIAL CONDITIONS}
\medskip
\centerline{\it 2.1 Method}
\medskip
The superiority
of the new TVD code over conventional hydrodynamic codes
(e.g., Cen 1992) and various tests of the code
have been presented
in Ryu \etal (1993);
essentially the shock capturing technique improves resolution
by approximately a factor of $2-3$ for
a given grid (i.e., nominal resolution).

The simulation reported
on in this paper did not include any atomic processes, i.e.,
no cooling or heating was added, except for the
adiabatic cooling due to the general expansion of the universe,
and ``heating" occurs only due to adiabatic compression
or to entropy generation at shock fronts.
For the hot gas, which we will discuss in this paper,
this approximation is valid,
since the cooling time exceeds the Hubble time
by a fair margin.
One can, after the fact, compute the emissivity with allowance
for line emission.
Doing this would increase the computed luminosity significantly
at and below 1keV but would
have little effect in the $2-10$keV band.
We will return to this important
matter in a subsequent paper devoted to
a comparison with observations.
Here we wish, primarily, to contrast open, $\Omega<1$,
and standard $\Omega=1$, variants  of the CDM scenario.

\medskip
\centerline{\it 2.2 Initial Conditions}
\medskip
We adopt a CDM+$\Lambda$ model
with the following parameters:
$n=1$,
$h=0.6$,
$\Omega=0.45$,
$\Omega_b=0.043$,
$\lambda=0.55$,
and
$\sigma_8=0.77$.
Note that the amplitude normalization of
the power spectrum is determined by
COBE observations
(Efstathiou, Bond, \& White 1993;
Kofman, Gnedin, \& Bahcall 1993)
parameterized by $\sigma_8$ to translate into conventional notation.
This chosen value of $\sigma_8$ corresponds
to a ``bias" of $b\equiv 1/\sigma_8=1.30$,
close to what is physically expected on this scale
(\cf Cen \& Ostriker 1992,1993) lending
further credibility to the adopted model.
Our box size is $85h^{-1}$Mpc
with $N=270^3$ cells and $135^3$ dark matter particles,
so our nominal resolution is $0.31h^{-1}$Mpc with
our real spatial resolution slightly worse than this.
The box size is determined by a compromise between two
considerations.
A larger box would allow longer waves and give us more of the rare,
high temperature, high luminosity clusters.
A smaller box (with a fixed $N$)
would allow us to resolve the cluster structure better.
The choice of $\Omega_b$ is
consistent with light element nucleosynthesis
(Walker \etal 1991).
The power spectrum transfer function is computed
using the method described in Cen, Gnedin, \& Ostriker (1993).
Gaussian initial conditions are used and the
same set of random numbers adopted as in KCOR.

\medskip
\centerline{3. RESULTS}
\medskip
The X-ray clusters in the simulation are identified as follows.
We first calculate the total X-ray luminosity due to thermal Bremsstrahlung
(assuming primeval composition and neglecting lines)
for each cell,
given the cell density and temperature,
assuming that hydrogen and helium are fully ionized (which is always
true in the regions like great clusters of galaxies).
The detailed formulae were presented in KCOR.
Note that in the following discussion all units
of length are given in co-moving, not metric, coordinates.

First we tag all cells having
total X-ray luminosity higher than $10^{38}$erg~s$^{-1}$.
This emissivity corresponds
to $3.2\times 10^{39}$erg/s/$h^{-3}$Mpc$^3$,
which is 3.2 times
the mean box emissivity at z=0,
and is 9.1 times the mean at z=3.
The number of X-ray bright cells defined
in this way is 47,826,
which comprises a fraction $2.4\times 10^{-3}$
of the box volume.
These
are selected as X-ray bright cells. Then we find the local maxima
(by comparing $L_{ff}$ of each X-ray bright cell with that
of 26 neighboring cells)
and identify them as the centers of the X-ray clusters.
Having defined the centers of the X-ray clusters, we go back
to the whole simulation box to define our X-ray clusters.
We analyze the simulation in the following
two different ways (which correspond to spheres of radius $0.5h^{-1}Mpc$
and $1.0h^{-1}$Mpc).
First, each cluster core
consists of $27$ cells (26 cells surrounding the central cell
plus the central cell).
These 27 cells are weighted so that the total volume of the cluster equals
the volume of a sphere of radius $0.5 h^{-1} Mpc$
as appropriate for observationally defined X-ray clusters
(see KCOR for details).
A similar algorithm is used for the $1.0h^{-1}$Mpc volumes.

Our ($85h^{-1}$Mpc)$^3$ box at $z=0$ contained
(0,0)
clusters with (total, core) luminosity brighter than $10^{45}$erg/s,
(1,0) brighter than $10^{44}$erg/s,
(10,5) brighter than $10^{43}$erg/s,
(66,40) brighter than $10^{42}$erg/s
and (257,174) brighter than $10^{41}$erg/s.
We explain the absence of clusters brighter than $10^{45}$erg/s
as simply due to the size of our box.
An additional factor of at least two in scale (and 8 in computer resources)
would be required to significantly improve
on the quoted results.
An immediate comparison to the KCOR results is possible if one notes
that for $L_x>10^{44}$erg/s,
the KCOR paper found
(8,1) clusters with
(total, core) luminosity,
higher than the specified limit compared to (1,0) in this work.
The difference is significant at this level of luminosity
and we will see below (Figure 1b and 4b)
that, while
the standard CDM model overproduces
bright X-ray clusters, this open model provides an adequate fit
to observations.
The fraction of the box in brightest cells
which provide
(50\%,90\%)
of the total box emissivity
is
 ($3.8\times 10^{-5}$,
 $8.8\times 10^{-4}$)

It is convenient to fit the luminosity
function to the three parameter Schechter function
$$\eqalignno{n(L)dL &= n_0 (L/L^{*})^{-\alpha} e^{-L/L^*} d(L/L^*)\quad .\quad
&(\the\eqtno )\cr}$$
\advance\eqtno by 1
Luminosity functions for cluster cores
were computed in four frequency bands:
total (bolometric) luminosity,
$0.3-3.5$keV, $0.5-4.5$keV, and $2-10$keV,
and are displayed in Figures 1a, 2a, 3a and 4a, respectively.
The results for entire clusters (emission from a $1 h^{-1}$Mpc sphere)
are presented for the same frequency bands in Figures 1b, 2b, 3b and 4b.
The figures show the domain of cluster properties in which
the observations and our computations overlap most:
$10^{40}$erg/s$\le L_x\le 10^{44}$erg/s and $0\le z\le 1$.
Also shown (cross-shaded areas) in Figures (1b) and (4b) are
the observations from Henry \& Arnaud (1991)
and from Henry (1992), respectively.
The comparison shown in (4b) is the more reliable one,
since line processes (omitted in our computation)
are unimportant in the 2-10keV band.

We see that the computed number densities of
bright clusters ($h^2 L\ge 10^{43}$erg/s)
are in accord with the observed ones while
in KCOR we found that those of the CDM model are above
the observed mean by a factor of about $5$.
We have computed approximate Schechter function fits to the results,
with the numerical parameters ($n_0,L^*,\alpha$)
as a function of redshift collected in Table (1),
and the simulated data extended to $z=5$.
Also in Table (1) we integrate over the cluster luminosity function,
using the Schechter fit,
$j_{cl}\equiv n_0L^*\Gamma(2-\alpha)$,
showing the results in the second-to-last column.
We give also, in the last column,
the total emissivity from the box as $j_{gas}$,
which includes the emission from lower density regions further from cluster
cores than $1.0h^{-1}$Mpc and also
from clusters whose central emissivity
is less than our cutoff value.
Note that $L_x^*$ and $L$ are in units of $10^{44}$erg/s;
$n_0$,
$n(L>10^{43})$
and $n(L>10^{44})$
are in units of $10^{-6}h^3$Mpc$^{-3}$;
$j_{cl}$ and $j_{gas}$ are in
units of $10^{40}$erg/s/h$^{-3}$Mpc$^3$, and $j_{cl}$ may be larger
than $j_{gas}$ due to the inaccuracy of the Schechter fit.

We estimate that $\alpha$ is well constrained ($\pm 0.03$)
and that the product $n_0 L^*$ is also fairly well constrained,
but individual values $(n_0,L^*)$ are poorly determined because
$L^*$ is dependent on the quite uncertain highest
luminosity clusters.
To estimate the purely statistical uncertainty,
we reanalyzed the $z=0$ data from the lower panel of Table 1a
(total, integrated X-ray cluster properties) looking separately
at two halves of the box.
The fractional differences ($|\Delta Q/Q|$) for $Q\equiv $
$(\alpha,L_x^*,kT_x,n_0,j_{cl},j_{gas})$
were found to be
$(0.0062,0.90,0.62,0.74,1.31,0.76)$
respectively.
The fact that even the integral $j$ varies
significantly between the two halves of the box
reminds us again of the ``cosmic variance".
Our sample volume is not large enough to give us a robust
estimate for the cosmic mean value of $j$.
Note that the variances of all the quantities (except for $\alpha$)
in this model are larger than those in the CDM model (KCOR),
presumably due to the fact that this CDM$+\Lambda$
model has relatively more
large-scale power than the CDM model,
indicating that a larger box is clearly desirable.
This, of course, is related
to the virtues of open models:
they typically have relatively more large-scale power
and thus match better a variety of
large-scale
structure observational constraints
(\cf, e.g., Efstathiou \etal 1993)
than do $\Omega=1$ models.

We see that the cluster cores, as we have defined them,
contain between $1/3$ and $1/2$ of the total X-ray emission
in the regions studied,
comparable to the results of KCOR.
The total cluster luminosity in the box
is typically $3/4$ of the X-ray emission
from the box, the same level was found in the CDM model (KCOR).
This fraction depends on energy,
as one can see by comparing Tables 1b and 4d.
Since most of the very high temperature gas is in the central
regions of clusters,
the fraction of the total emission from identifiable
clusters in the $2-10$keV region
approaches unity.
For the total luminosity the Schechter $\alpha$ parameter is approximately
$1.55$ (as compared to 1.50 in KCOR)
with little evolution, and $\alpha$, for the few keV
bands is typically slightly flatter at $1.4$.
For $\alpha < 2$, as we have noted,
most of the luminosity arises from bright clusters.
Observationally $\alpha=1.9-2.0$;
so this
model provides a marginally better fit than does
the standard $\Omega=1$, CDM model.
Allowance for line emission processes
would increase the luminosity of the lower luminosity (low temperature)
clusters more than the high luminosity (high temperature) clusters,
still further improving the fit.
Without a careful treatment of line emission,
it would be premature
to say if the fit to the slope of the luminosity function, as
represented by $\alpha$
in this model,
is adequate.

The number density of bright clusters
peaks at intermediate redshift,
and the typical luminosity is (for small redshift)
relatively constant, so there is
a peak emissivity at approximately
$z=0.2-0.5$ for the several keV bands.
Thus, crudely speaking,
in this model one expects weak ``positive" evolution until nearly
$z=0.5$ and then
negative evolution thereafter.
The peak occurs at significantly lower
redshift in the open model than in the standard CDM model.
Figure (5a) shows the comparison between the two models
for the number density (per unit comoving volume)
of clusters with rest frame luminosity
$>10^{43}$erg/s.
Figure (5b) shows
the number density (per unit comoving volume)
of clusters with rest frame luminosity (integrated over the entire
frequency range)
$>10^{43}$erg/s
for five different models.
We see that it is the overall normalization
on the relevant scale ($\sigma_8$) rather
than $\Omega$
which is the dominant factor on the evolutionary
behavior of the bright cluster number density.
Physically, the rapid evolution
occurs at an epoch when
most of the bright clusters are collapsing.
After that, the evolution still goes on due to
processes such as merging, but it is relatively mild.
Comparing the COBE-normalized
standard CDM and COBE-normalized CDM$+\Lambda$ models,
negative evolution begins earlier and is stronger
for the open model.
Allowance for cosmological effects would make
both models evolve more negatively,
but these effects would be
exaggerated in the CDM$+\Lambda$ case due
to the rapid increase in luminosity distance
(with redshift) in
cosmologies with a significant cosmological constant.
The different evolutionary path
of the cluster luminosity
function is the most
significant difference
we have found
between open and $\Omega=1$ models.
For example,
Figure (5b) shows more than an
order of magnitude decline
in the number density
of clusters
with 0.5-4.5keV
luminosity $>10^{43}$erg/s
between redshift zero
and two for the open model,
but it shows a slight increase for the standard CDM $\Omega=1$ model.

We believe that the peak in emission seen at moderate redshift is real.
The reasons for this are discussed in KCOR.
The approximate Press-Schechter formalism, which does not
allow for a variety of effects, cannot easily
mimic the full non-equilibrium hydrodynamic treatment.

While the integrated X-ray emissivity evolves fairly slowly over the period
surveyed in Table (1a) with a
flat maximum between $z=0.2$ and $z=0.3$,
$L^*$ and $n_0$ tend to evolve more rapidly and in opposite directions.
The primary difference which we find between this model and
the CDM model (KCOR) is the evolution of the bright end
X-ray clusters.
In the $\Omega=1$ CDM model (KCOR) we find that
$L^*$ is a monotonic decreasing function of redshift,
while in this model we find $L^*$ peaks at $z\sim 0.3$.
The reason for this difference is the dramatically
different behaviors of the universal expansion at later
times in the two models.
We expect that in a lower $\Omega$ model, $L^*$
will clearly peak at an even higher redshift.
The evolution of $L^*$ should be observable and
provide a powerful discrimenant among cosmic models.

To highlight the negative evolution of the
bright end
of the luminosity function,
we listed in the fifth and sixth columns
of Tables 1b-1d (sixth and seventh columns of Table 1a)
the comoving density of clusters having luminosity greater than
$10^{43}$erg/s and
$10^{44}$erg/s.
For reasons stated earlier (based on our limited box size),
we use the Schechter fit rather than
direct counts to compute these columns.
Comparing columns 4 and 6 (5 and 7 of Table 1a),
we see that, although the total number
density $n_0$
of clusters increases with redshift
(until $z\approx 1-2$),
the number density
of the highest luminosity ($L_x>10^{44}$erg/s)
clusters decreases for $z>0.5$.
This is presumably one of
the effects leading to the observational
appearance of ``negative evolution".
Statistical fluctuations in our results are
still quite significant due to
the limited box size, especially for $n(L_x>10^{44}$erg/s$)$.

Redshift effects strongly exaggerate this
tendency to observe
negative evolution,
since higher redshift clusters
tended to have lower temperatures (\cf column 4
of Table 1a and Figures 6, 13),
and both effects will reduce the energy observed
by satellites measuring the X-ray flux in high energy bands.
Note that the negative evolution in the density
of clusters with $L>10^{44}$erg/s is more
and more steep in Tables (a$\rightarrow$ d)
as one looks at higher energy bands,
although statistical fluctuations are large.

The emission-weighted temperature,
$T_x$, of each cluster is
calculated and the distributions
are shown in Figure (6).
The arrow in each panel indicates the average
cluster temperature (weighted by luminosity) at the given epoch.
Also shown in Figure (6b) is
the observed temperature function from Henry \& Arnaud (1991)
as the cross-shaded area.
Note that the computed temperature function is somewhat lower
than the observed one, but we think that the limited size of
our simulation box caused an omission from
the computation of the highest temperature clusters;
a bigger box is needed before we can
have definite conclusions on this.
We see that at all epochs the coolest clusters
dominate the statistics (the turnover
at low $T_x$ is presumably caused by our
definition of minimun cell luminosity to
constitute an X-ray bright cell),
but the mean is determined by the high mass, high luminosity,
high temperature end of the distribution.
The mean temperatures, indicated by arrows,
are included in column 3 of Table (1a).
We will return to the issue of temperature evolution later.
Looking ahead to Figure (8a), we see the strong correlation
found between $T_x$ and total luminosity
(clusters are shown at $z=0$).

Now let us turn again to the total
cluster luminosity (vs. core luminosity)
as shown in the lower panels of the tables and
in Figures 1b-4b,
the quantity normally measured by
satellite observations.
The ratio of $(j_{cl,tot}/j_{cl,core})$ is
near $2.0\pm 0.1$ for the redshift range $0<z<1$;
a similar value was found in KCOR.

We can also roughly estimate the effective radii of the clusters
by assuming that the emission has a profile
$$\eqalignno{j &= {j_0\over [1+(r/r_x)^2]^2}\quad \quad &(\the\eqtno )\cr}$$
\advance\eqtno by 1
and determining, from the ratio of the luminosity
(integrated over frequency)
of the central cell to the total cluster luminosity,
the value of $r_x$ which would produce this ratio.
We show in Figure (7) the radii
determined in this fashion.
The peaks seen in the panels
of Figure (7) are, of course, artificially
induced by our
cell size of $0.31h^{-1}$Mpc,
but the distribution to larger radii should be reasonably accurate.
Arrows indicate the luminosity-weighted average values.
Since brighter clusters tend to be resolved,
these numbers should be reliable,
but the arrows are uncomfortably close to the
peaks of the curves,
indicating that the most luminous clusters may be somewhat unresolved.
We see a weak trend of increasing size with increasing
time,
which is in the theoretically anticipated direction.
Longer wavelengths became nonlinear later,
producing larger clusters, and smaller clusters
merge to produce larger clusters with increasing time.

Now in Figures (8a,b)
we show the scatter plots of ($T_X,r_x$) vs. $L_{x}$
(integrated over frequency).
We see that there is a clear correlation between $L_x$ and $T_x$.
But we do not see any strong correlation
between $L_x$ and $r_x$.
The best fit lines (dashed)
indicate a slope of ($0.36\pm 0.01 ,0.32\pm 0.01$)
for (core, total) cluster region.
In KCOR we found ($0.394\pm 0.001 ,0.375\pm 0.001$) for the CDM model.
The observed correlation [\cf Figure (8a) shown as cross-shaded area]
between $\bar T_x$ and total cluster
luminosity is
$\log_{10} \bar T_x ({\rm keV}) = \log_{10} (4.2^{+1.0}_{-0.8}) + (0.265\pm
0.035)\log_{10} (h^2L_{44})$,
according to Henry \& Arnaud (1991).
In the region where a comparison can be made,
the agreement with observations is good.

Next, we address temperature variations within clusters.
Given our limited resolution, there is little that can be
accurately determined on this issue from our simulations,
but we are able
to compare the central cell (Volume$=3.1\times 10^{-2}h^{-1}$Mpc$^3$)
with the surrounding cells ($3^3-1^3$)
(volume of size $5.9\times 10^{-1}h^{-3}$Mpc$^3$)
and the cells surrounding these cells ($5^3-3^3$, vol$=3.0 h^{-3}$Mpc$^3$).
We define the ratio of the inner cell to
the next cube
as $T_c/T_{shell}$ (volume-weighted average)
and show the scatter diagram in Figure (9).
No trend is seen with luminosity, and
the median value, indicated by the dashed line, is $1.4$.
The cluster gas deviates significantly from
isothermality,
with a $5\%-10\%$ temperature
decline typically found by a radius of
$0.4-0.5h^{-1}$Mpc but a sharp
fall-off is indicated for radii $1h^{-1}$Mpc.
In Figure (10) we compare the (luminosity-weighted) temperatures found
in the three regions noted above,
normalized to the temperature in the central cell.
The large dispersion is indicated by the error bar.
Figures (9) and (10)
show no significant differences from the
analogous figures in the $\Omega=1$ case (KCOR).

We show in Fig. 11 and 12 the evolution of cluster core radius
(in metric, not comoving, units)
and temperature (both luminosity-weighted) as open circles.
Solid dots are from the KCOR $\Omega=1$ computation.
Also shown are the fits
analytically predicted by Kaiser (1986)
$R_x\propto (1+z)^{-2}$ and $T_x\propto (1+z)^{-1}$
for $\Omega=1$, CDM model (solid line) from KCOR
and
this CDM$+\Lambda$ model (dotted line)
(note that Kaiser's prediction is valid
for $\Omega=1$ models only, but is
shown also for the CDM$+\Lambda$ model for comparison purposes).
Examination of these figures indicates
the radius changes in a way qualitatively
similar to that found for standard CDM
with an $\sim 40\%$ decline in radius from redshift zero
to a look-back of redshift $z=0.5$.
But there is an overall difference in that the radii are smaller in
the CDM$+\Lambda$ model.
The difference, while small ($-23\%$),
is significant, especially since the lower bound produced
by our finite numerical resolution is less than
a factor of two below the mean value.
Since real, observed clusters are smaller than
the computed clusters, this change
is also in the desired direction.
The differences between the two models
are dramatic with regard to temperature.
First, as noted, the mean temperature is a factor of 3.5 lower
in the open model and the evolution is far slower
($-33\%$ to $z=0.5$ rather than $-48\%$ for $\Omega=1$).

Thus we see that
the differences
between the two models
at low redshift
are significant and very important,
due to
freeze-out at low redshift in open models.
This temperature evolution
should be detectable even with a relatively ``soft" X-ray instrument
such as ROSAT.
It will be able to discriminate between
$\Omega=1$ models
and $\Omega <1$ models.
Furthermore, detailed and direct
numerical simulations combined with more observations
might provide a way to constrain $\Omega$, or $\lambda$,
or combinations of $\Omega$ and $\lambda$.

Figures (13) and (14) show the evolution in the background radiation
field in two additional ways.
The first shows what a comoving observer would
have measured at various redshifts.
But below 1keV the results are unreliable
because of both the omission of line emission and the omission
of IGM absorption.
The second, Figure (14), shows the fractional contribution
to the background seen by an observer at $z=0$ in several bands
that were produced at various epochs (in integral form).
The important point to note is
that most of the X-ray background (especially
in the harder bands)
that we see locally were produced at relatively
recent ($z\le 0.5$) epochs.
This is a consequence of many things,
prominant among them are the redshift factors that dilute
the observable effects of emission at high redshift.
But there is a major difference between
the two models.
In this open model, one-half of the $2-10$keV
brightness of the sky comes from redshift less than $z=0.3$,
whereas, in the $\Omega=1$ case (KCOR),
the median point is much closer, at redshift $z=0.2$.

Finally, let us take a slightly different route
to address the issue of bias of gas relative to the mass:
does the gas in dense regions, like clusters of galaxies,
fairly represent the underlying mass,
or is it ``biased" or anti-biased?
This is a question with great cosmological significance.
If we know the ratio of gas (+ galaxies) to total matter
in the clusters by direct observations,
and we know, from light element nucleosynthesis the global
baryon density,
then we can divide the second number by the first to obtain the global
matter density and to compare with the
cherished critical density.
This old line of argument has been carefully
re-examined recently by
White (1992) and also reanalyzed by
Babul \& Katz (1993) and others.

The argument depends on knowing whether or not
$\rho_b/\rho_{tot}$ varies significantly from place to place and,
in particular, whether this quantity
will be found near its average value in the high density regions,
where it can most easily be measured.

Our possibly counter-intuitive results are shown in Figure (15),
where we plot the ratio $(\rho_{IGM}/\rho_{tot})$
vs. $(\rho_{tot}/\langle\rho_{tot}\rangle)$
smoothed by a gaussian of radius $1h^{-1}$Mpc.
At any value for
the total density, there is a wide range of possible
values of $\rho_{IGM}$, but
the high density regions actually
have a {\it lower} than average ratio
of baryons to total mass.
We see that, in the high density clusters,
where $\rho_{tot}/<\rho_{tot}>$ approaches $10^3$,
the gas is under represented by a factor of $1.1$.
(In KCOR we found a factor of $1.7$ for the CDM model.)
Combining this factor with its global mean,
we obtain that $({\rho_{gas}\over\rho_{tot}})_{cl}=0.083\pm 0.007$,
which is consistent
with observations
[$({\rho_{gas}\over\rho_{tot}})_{obs}=0.097\pm 0.019$
for the Coma cluster for the given value of $h$, \cf White 1991].

We can use the comparison with KCOR to address
the question of whether or not
the small anti-bias found is real or due merely
to numerical errors spreading out the gas density more than it
does the dark matter density.
In this simulation the core radii of the clusters
are smaller on average than in the $\Omega=1$ case,
putting them in better agreement with observations
but, as noted,
closer to our grid spacing.
Thus numerical diffusion in this case should be a larger (relative)
effect
here than in KCOR, and
the anti-bias, if it were due to numerical diffusion,
would be larger.
But it is smaller, indicating that this is not the case.
Our best guess is that the cause of the
anti-bias is related to outward propagating
shocks in the transient formation phase.
Since such time-dependent effects will be less in open
models, the anti-bias should be less here -- as it is.

Now, we treat the same problem in a somewhat different way concentrating
on the identified clusters.
For each simulated X-ray cluster,
we compute the gas
mass and total mass within a sphere of radius $1h^{-1}$Mpc (centered
at the X-ray cluster center) and in Figure (16) we show
the ratio (open circles)
of the two masses within this sphere as a function
of density of the sphere relative to the mean.
The dotted line is the best log-log straight line fit for the
open circles, and the solid line is the fit weighted by the luminosity
of each cluster.
Also shown, as the dashed line, is the global mean of the ratio.
Presenting the same information in an alternative way,
we show in Figure (17) the histogram of the
ratio,
both number-weighted (thin solid histogram)
and luminosity-weighted (thin dotted histogram)
in the CDM$+\Lambda$ model.
The heavy solid histogram (arbitrarily normalized to
have a similar peak height)
indicating the observational
situation is adapted
from Jones \& Forman (1992).
We see that there is a trend that poor clusters
are relatively gas poor, which is consistent
with observations (\cf, e.g., Jones \& Forman 1992).
The luminosity-weighted fit is closer to (but still below) the
global mean with an anti-bias in the range $0.85-0.92$.
We see that the median of the computed ratio is in agreement
with the median of the observed ratio,
whereas in the CDM $\Omega=1$ model the computed ratio
is lower by a factor of 2-3.
Improving the observations will probably narrow the heavy histogram,
and increasing the dynamical range of our simulation box
will widen the thin histograms.
Both expected improvements should make the agreement better.

\medskip
\centerline{4. CONCLUSIONS}
\medskip
In Figures (1-5) we show
the evolution of the cluster
luminosity
function expected in the CDM$+\Lambda$ model.
In the range of parameters
where there is greatest overlap between observed and computed quantities
($0\le z\le 1$, $10^{40}$erg/s$\le L_x\le 10^{44}$erg/s),
little evolution is seen (for comoving observers) in any
of the computed bands aside from a decline
in the number of brightest sources
(somewhat uncertain due to our limited box size)
and a modest increase (by about a factor of two)
in the luminosity
function for fainter objects.
A similar behavior was also found in the CDM model (KCOR).
But the likely explanation for this is different
from that in the CDM case.
While in the standard CDM model,
it is largely coincidental and due to the
balancing of two effects:
new breaking waves increase the luminosity density
but mergers decrease it.
In this CDM$+\Lambda$ model it is primarily
due to the late time
freeze-out of formed clusters.
The effect is seen most clearly in Figure (5),
where
we look only at the evolution of the brightest
clusters,
and see an earlier decline (with increasing redshift)
of the open model than the $\Omega=1$ models.
There is an order of magnitude decline
in the number density of clusters having
$L_x>10^{43}$erg/s
(in the harder energy bands)
in the redshift interval
$z=0\rightarrow 2$ for this model
but an increase in the standard $\Omega=1$ CDM model.
The strong negative evolution found in this paper
in an open model
would of course
be greatly
enhanced
for observers
(using fixed bands and intensity limits)
by cosmological effects which
are especially strong in models with a cosmological
constant (Carroll, Press, \& Turner 1992)

Figures 6 and 7 show
rates of change in other quantities,
the temperatures and clusters radii.
These important trends are summarized in Figures 11
and 12 where we see a factor of 2-3 decline
in both these quantities by redshift $1$.
But more interesting is the nearly constant mean temperature
(luminosity-weighted)
in the redshift range $z=0\rightarrow 0.3$ in the CDM$+\Lambda$
while in the CDM model we see a sharp decrease in the temperature
from $z=0.0$ to $z=0.3$.
Equivalently put, $L^*$ of Schechter fits to the computed luminosity
functions peaks near $z=0.3$ in the CDM$+\Lambda$ model
but it increases monotonically until $z=0$ in the CDM model.
Both trends should be detectable even with
a relatively ``soft" X-ray instrument
such as ROSAT.
They might provide powerful tests for $\Omega=1$ models
and $\Omega<1$ models.
Also the actual values of the cluster
temperature are much lower in the open model
and in better agreement with observations.

This CDM$+\Lambda$ model provides an adequate fit to
the observed bright X-ray cluster luminosity function
and X-ray background; however,
we found that there would be too many bright X-ray clusters produced
and too much integrated background X-ray
intensity in the COBE-normalized standard CDM model.

We find a slight anti-bias ($\sim 10\%$) of gas relative to
the mass in dense regions like the clusters of galaxies,
but the model is self-consistent in the sense
that
the computed $({\rho_{gas}\over\rho_{tot}})_{cl}$($=0.083\pm 0.007$)
is consistent with observations
[$({\rho_{gas}\over\rho_{tot}})_{obs}=0.097\pm 0.019$
for the Coma cluster for the given value of $h$, \cf White 1991],
whereas there was a gross inconsistency
in the $\Omega=1$ case.

In sum, the model is significantly different
from that obtained from the standard $\Omega=1$,
CDM simulation, and, with regard to all measurable
quantities that we have compared to observations
[$N(L_x)$, $\langle R_x\rangle$, $\langle T_x\rangle$,
$dln T_x/dln L_x$,
$\rho_{gas}/\rho_{tot}$],
it is not only a better fit to observations than standard $\Omega=1$,
CDM model,
but also is an adequate representation of them.
The predicted evolutionary
differences between open and $\Omega=1$ models
are sufficiently great
to allow definitive tests
by current or planned X-ray satellite observations.

\medskip
\medskip
We are happy to acknowledge
support from NASA grant
NAGW-2448, NSF grant AST91-08103 and the NSF HPCC grant ASC93-18185.
We would like to especailly thank R. Reddy for his valuable
help and patient effort to optimize the code on the Cray-90 supercomputer.
It is a pleasure to acknowledge
the Pittsburgh Supercomputer Center for allowing
us to use their Cray-90 supercomputer and the help
of Koushik Ghosh from Cray Reseach on the FFT routines.
Discussions with Nick Gnedin, Patrick Henry and Simon White
are gratefully acknowledged.

\vfill\eject

\centerline{REFERENCES}
\refset
Bryan, G.L., Cen, R.Y., Norman, M.L., Ostriker, J.P., \& Stone, J.M. 1994, ApJ,
 in press
\smallskip
\refset
Carroll, S.M., Press, W.H., \& Turner, E.L. 1992, ARAA, 30, 499
\smallskip
\refset
Cen, R.Y., 1992, ApJS, 78, 341
\smallskip
\refset
Cen, R.Y., \& Ostriker, J.P, 1992, ApJ(Letters), 399, L113
\smallskip
\refset
Cen, R.Y., \& Ostriker, J.P, 1993, ApJ, 417, 415
\smallskip
\refset
Cen, R.Y., Gnedin, N.Y., \& Ostriker, J.P, 1993, ApJ, 417, 387
\smallskip
\refset
Efstathiou, G. 1992, NAS Colloqium on Physical Cosmology
\smallskip
\refset
Henry, J.P. 1992, in ``Clusters and Superclusters of Galaxies",
ed. A.C. Fabian (Kluwer Academic Publisher. Printed in the Netherland),
p311
\smallskip
\refset
Henry, J.P., \& Arnaud, K.A. 1991, ApJ, 372, 410
\smallskip
\refset
Jones, C., \& Forman, W. 1992 in ``Clusters and Superclusters of Galaxies"
(Kluwer Publishers: Dordrecht), ed. A.C. Fabian, p49
\smallskip
\refset
Kaiser, N. 1986, MNRAS, 222, 323
\smallskip
\refset
Kang, H., Cen, R.Y., Ostriker, J.P., \& Ryu, D. 1994, ApJ, in press (KCOR)
\smallskip
\refset
Kofman, L.A, Gnedin, N.Y., \& Bahcall, N.A. 1993, ApJ, 413 1
\smallskip
\refset
Ostriker, J.P. 1993, ARAA, 31, 689
\smallskip
\refset
Ryu, D., Ostriker, J.P., Kang, H., \& Cen, R.Y. 1993, ApJ, 414, 1
\smallskip
\refset
Walker, T.P., Steigman, G., Schramm, D.N., Olive, K.A., and Kang, H.S.,
   1990, ApJ, 376, 51
\smallskip
\refset
White, S.D.M 1992 in ``Clusters and Superclusters of Galaxies"
(Kluwer Publishers: Dordrecht), ed. A.C. Fabian, p17
\smallskip
\refset
Wu, X., Hamilton, T., Helfand, D.J., \& Wang, Q. 1991, ApJ, 379, 564
\smallskip
\refset
\vfill\eject

\centerline{FIGURE CAPTION}
\medskip

\item{Fig. 1--}
Figure (1a):
the X-ray cluster bremsstrahlung luminosity (from central
$<0.5h^{-1}Mpc$ regions) function
integrated over the
whole frequency range at five different redshifts
$z=(0, 0.2, 0.5, 0.7, 1.0)$.
Figure (1b):
the X-ray cluster bremsstrahlung luminosity (from
$<1.0h^{-1}$Mpc region) function
integrated over the
whole frequency range (filled dots) at the
same five different redshifts.
The cross-shaded area shows the observations
(Henry \& Arnaud 1991,
$\{3.1^{+4.5}_{-1.8}\times 10^{-6} h^3 {\rm Mpc}^{-3} h^2 [L_{44}({\rm
bol})]^{-1}\}
\times [h^2 L_{44}({\rm bol})]^{-1.85\pm 0.4}$).

\item{Fig. 2--}
Same as Figure (1) but for the luminosities integrated
over $0.3-3.5$keV frequency bin.

\item{Fig. 3--}
Same as Figure (1) but for the luminosities integrated
over $0.5-4.5$keV frequency bin.

\item{Fig. 4--}
Same as Figure (1) but for the luminosities integrated
over $2-10$keV frequency bin.
The cross-shaded area in (4b) indicates observations (Henry 1992).

\item{Fig. 5--}
Figure (5a) shows the comparison between the two models
for the number density (per unit comoving volume)
of clusters with rest frame luminosity
$>10^{43}$erg/s for three X-ray bands.
Figure (5b) shows
the number density (per unit comoving volume)
of clusters with rest frame luminosity (integrated over the entire
frequency range)
$>10^{43}$erg/s
for five different models.

\item{Fig. 6--}
The X-ray cluster temperature ($T_x$, emission-weighted temperature)
function at six different redshifts $z=(0,0.5,1,2,3,5)$.
Arrows indicate the luminosity-weighted average temperature $\bar T_x$
at each epoch.
In the first ($z=0$) panel in (5b) the cross-shaded area
is the observed temperature function from
Henry \& Arnaud (1991)
[$(1.8^{+0.8}_{-0.5}\times 10^{-3} h^3 {\rm Mpc}^{-3} keV^{-1})
(kT)^{-4.7\pm 0.5}$].

\item{Fig. 7--}
The X-ray cluster effective radius ($r_x$)
distribution [\cf equation (2)].
Arrows indicate the luminosity-weighted effective radius at each epoch.

\item{Fig. 8--}
Figure (8a) shows the scatter plot of $T_x$ vs $L_{x}$ at $z=0$.
The cross-shaded area line indicates the observations
of Henry \& Arnaud (1991).
The dashed line is the best fit of the simulation results.
Figure (8b) shows the scatter plot of $r_x$ vs $L_{x}$ at $z=0$.

\item{Fig. 9--}
The ratio of the central cell temperature to the temperature
of its surrounding shell ($\sim$ one cell thick) as a function
of $L_{tot}$.

\item{Fig. 10--}
We compare the (luminosity-weighted) temperatures found
in the three regions (central cell, the shell surrounding the central
and the next outer shell)
and normalized to the temperature in the central cell.
Departure from isothermality increase
significantly for $hr>0.5$Mpc.
Note the errorbars are $1 \sigma$ variance.

\item{Fig. 11--}
The average cluster core radii in physical units as a function of redshift
for clusters with luminosity in the $0.5-4.5$keV band greater than
$10^{43}$erg/s for CDM$+\Lambda$ model (open circles, this paper)
and the standard $\Omega=1$, CDM model (solid dots, KCOR).
The best fit evolutions of the form $T_x\propto (1+z)^{-2}$
are shown as a solid curve for the CDM model
and a dotted curve for the CDM$+\Lambda$ model.

\item{Fig. 12--}
The average cluster temperature as a function of redshift
for clusters with luminosity in the $0.5-4.5$keV band greater than
$10^{43}$erg/s for CDM$+\Lambda$ model (open circles, this paper)
and the standard $\Omega=1$, CDM model (solid dots, KCOR).
The best fit evolutions of the form $T_x\propto (1+z)^{-1}$
are shown as a solid curve for the CDM model
and a dotted curve for the CDM$+\Lambda$ model.
Temperatures in the CDM$+\Lambda$ model
are lower and evolution less than
in the standard $\Omega=1$, CDM model,
and tend to freeze out at lower redshift while
we see a dramatic increase in the CDM model approaching $z=0$.

\item{Fig. 13--}
The mean radiation intensity at six epochs,
$z=5$ (solid line),
$z=3$ (dotted line),
$z=2$ (short, dashed line),
$z=1$ (long, dashed line)
$z=0.5$ (dotted-short-dashed line) and
$z=0$ (dotted-long-dashed line).
The box in the middle shows the obseravtional data by Wu \etal (1991).
Neither line absorption nor emission has been allowed for
in this figure.

\item{Fig. 14--}
The distribution functions of four presently observed X-ray bands
as a function of redshift (in integral form).

\item{Fig. 15--}
The ratio $\rho_{IGM}/\rho_{tot}$ as a function
of $\rho_{tot}/<\rho_{tot}>$.
Results are smoothed by a gaussian window of
radius $1h^{-1}$Mpc.
The global mean value of $\rho_{IGM}/\rho_{tot}$
is shown by the dashed line.
Note that in the highest density regions the gas is under-represented,
``anti-biased", by a factor of about $1.1$
(which is less than 1.7, found for $\Omega=1$ CDM model in KCOR).

\item{Fig. 16--}
The ratio $\rho_{gas}/\rho_{tot}$ as a function
of $\rho_{tot}/<\rho_{tot}>$ within a radius
of $1h^{-1}$Mpc for each identified cluster (open circles).
The dotted line is the best log-log straight line fit for the
open circles and the solid line the fit weighted by the luminosity
of each cluster.
Also shown as the dashed line is the global mean of the ratio.
We see that there is a trend that poor clusters
are relatively gas poor.

\item{Fig. 17--}
The histogram of the
ratio both number-weighted (thin solid histogram)
and luminosity-weighted (thin dotted histogram)
in the CDM$+\Lambda$ model.
The heavy solid histogram indicating the observational
situation is adapted
from Jones \& Forman (1992).
We see that there is a trend that poor clusters
are relatively gas poor, which is consistent
with observations (\cf, e.g., Jones \& Forman 1992).
The luminosity-weighted fit is closer to (but still below) the
global mean,
with an anti-bias in the range $0.85-0.92$.
We see that the median of the computed ratio is in agreement
with the median of the observed ratio, whereas
in the CDM $\Omega=1$ model the computed ratio
is lower by a factor of 2-3.
Improving the observations will probably narrow the heavy histogram,
and increasing the dynamical range of our simulation box
will widen the thin histograms.
Both expected improvements should make the agreement better.

\vfill\eject

\centerline {{\bf Table 1a.} Parameters of Schechter fits for the
 X-ray cluster luminosity function}
\vskip -0.3cm
\centerline {integrated over the entire frequency range.}
\vskip -0.3cm
\centerline {X-ray Cluster Core Luminosity ($<0.5h^{-1}$Mpc)}
\begintable
\hfill \quad $z$ \qquad\hfill|
\hfill \quad $\alpha$ \qquad\hfill|
\hfill \quad $L_x^*$ ($10^{44}$) \qquad\hfill|
\hfill \quad $k\bar T_x$ (keV) \qquad\hfill|
\hfill \quad $n_0$ \qquad\hfill|
\hfill \quad $n(L>10^{43})$ \qquad\hfill|
\hfill \quad $n(L>10^{44})$ \qquad\hfill|
\hfill \quad $j_{cl}$ \qquad\hfill|
\hfill \quad $j_{gas}$\qquad\hfill\cr
\hfill \quad 0 \qquad \hfill|
\hfill \quad 1.52 \qquad \hfill|
\hfill \quad 0.54 \qquad \hfill|
\hfill \quad 1.79 \qquad \hfill|
\hfill \quad 4.52 \qquad \hfill|
\hfill \quad 3.9 \qquad \hfill|
\hfill \quad $7.4\times 10^{-2}$ \qquad \hfill|
\hfill \quad 0.045 \qquad \hfill|
\hfill \quad 0.10 \qquad \hfill\cr
\hfill \quad 0.2 \qquad \hfill|
\hfill \quad 1.60 \qquad \hfill|
\hfill \quad 1.16 \qquad \hfill|
\hfill \quad 1.51 \qquad \hfill|
\hfill \quad 2.39 \qquad \hfill|
\hfill \quad 4.7 \qquad \hfill|
\hfill \quad 0.24 \qquad \hfill|
\hfill \quad 0.061 \qquad \hfill|
\hfill \quad 0.12 \qquad \hfill\cr
\hfill \quad 0.5 \qquad \hfill|
\hfill \quad 1.55 \qquad \hfill|
\hfill \quad 0.27 \qquad \hfill|
\hfill \quad 1.05 \qquad \hfill|
\hfill \quad 10.7 \qquad \hfill|
\hfill \quad 4.2 \qquad \hfill|
\hfill \quad $1.1\times 10^{-2}$ \qquad \hfill|
\hfill \quad 0.053 \qquad \hfill|
\hfill \quad 0.14 \qquad \hfill\cr
\hfill \quad 0.7 \qquad \hfill|
\hfill \quad 1.54 \qquad \hfill|
\hfill \quad 0.15 \qquad \hfill|
\hfill \quad 0.84 \qquad \hfill|
\hfill \quad 19.1 \qquad \hfill|
\hfill \quad 3.1 \qquad \hfill|
\hfill \quad $4.8\times 10^{-4}$ \qquad \hfill|
\hfill \quad 0.055 \qquad \hfill|
\hfill \quad 0.13 \qquad \hfill\cr
\hfill \quad 1 \qquad \hfill|
\hfill \quad 1.53 \qquad \hfill|
\hfill \quad 0.11 \qquad \hfill|
\hfill \quad 0.64 \qquad \hfill|
\hfill \quad 27.7 \qquad \hfill|
\hfill \quad 2.6 \qquad \hfill|
\hfill \quad $4.1\times 10^{-5}$ \qquad \hfill|
\hfill \quad 0.057 \qquad \hfill|
\hfill \quad 0.12 \qquad \hfill\cr
\hfill \quad 2 \qquad \hfill|
\hfill \quad 1.58 \qquad \hfill|
\hfill \quad 0.043 \qquad \hfill|
\hfill \quad 0.31 \qquad \hfill|
\hfill \quad 37.9 \qquad \hfill|
\hfill \quad 0.28 \qquad \hfill|
\hfill \quad $8.8\times 10^{-12}$ \qquad \hfill|
\hfill \quad 0.034 \qquad \hfill|
\hfill \quad 0.077 \qquad \hfill\cr
\hfill \quad 3 \qquad \hfill|
\hfill \quad 1.93 \qquad \hfill|
\hfill \quad 0.22 \qquad \hfill|
\hfill \quad 0.14 \qquad \hfill|
\hfill \quad 1.14 \qquad \hfill|
\hfill \quad 0.38 \qquad \hfill|
\hfill \quad $2.1\times 10^{-4}$ \qquad \hfill|
\hfill \quad 0.035 \qquad \hfill|
\hfill \quad 0.035 \qquad \hfill\cr
\hfill \quad 5 \qquad \hfill|
\hfill \quad 1.90 \qquad \hfill|
\hfill \quad 0.11 \qquad \hfill|
\hfill \quad 0.054 \qquad \hfill|
\hfill \quad 0.37 \qquad \hfill|
\hfill \quad $3.1\times 10^{-2}$ \qquad \hfill|
\hfill \quad $2.3\times 10^{-7}$ \qquad \hfill|
\hfill \quad 0.0039 \qquad \hfill|
\hfill \quad 0.0086 \qquad \hfill
\endtable
\centerline {X-ray Cluster Total Luminosity ($<1h^{-1}$Mpc)}
\begintable
\hfill \quad $z$ \qquad\hfill|
\hfill \quad $\alpha$ \qquad\hfill|
\hfill \quad $L_x^*$ ($10^{44}$) \qquad\hfill|
\hfill \quad $k\bar T_x$ (keV) \qquad\hfill|
\hfill \quad $n_0$ \qquad\hfill|
\hfill \quad $n(L>10^{43})$ \qquad\hfill|
\hfill \quad $n(L>10^{44})$ \qquad\hfill|
\hfill \quad $j_{cl}$ \qquad\hfill|
\hfill \quad $j_{gas}$\qquad\hfill\cr
\hfill \quad 0 \qquad \hfill|
\hfill \quad 1.56 \qquad \hfill|
\hfill \quad 1.74 \qquad \hfill|
\hfill \quad 1.55 \qquad \hfill|
\hfill \quad 2.56 \qquad \hfill|
\hfill \quad 6.6 \qquad \hfill|
\hfill \quad 0.54 \qquad \hfill|
\hfill \quad 0.090 \qquad \hfill|
\hfill \quad 0.10 \qquad \hfill\cr
\hfill \quad 0.2 \qquad \hfill|
\hfill \quad 1.56 \qquad \hfill|
\hfill \quad 1.86 \qquad \hfill|
\hfill \quad 1.29 \qquad \hfill|
\hfill \quad 3.14 \qquad \hfill|
\hfill \quad 8.5 \qquad \hfill|
\hfill \quad 0.73 \qquad \hfill|
\hfill \quad 0.12 \qquad \hfill|
\hfill \quad 0.12 \qquad \hfill\cr
\hfill \quad 0.5 \qquad \hfill|
\hfill \quad 1.54 \qquad \hfill|
\hfill \quad 0.59 \qquad \hfill|
\hfill \quad 0.91 \qquad \hfill|
\hfill \quad 9.64 \qquad \hfill|
\hfill \quad 9.2 \qquad \hfill|
\hfill \quad 0.20 \qquad \hfill|
\hfill \quad 0.11 \qquad \hfill|
\hfill \quad 0.14 \qquad \hfill\cr
\hfill \quad 0.7 \qquad \hfill|
\hfill \quad 1.50 \qquad \hfill|
\hfill \quad 0.29 \qquad \hfill|
\hfill \quad 0.72 \qquad \hfill|
\hfill \quad 20.4 \qquad \hfill|
\hfill \quad 8.6 \qquad \hfill|
\hfill \quad $3.2\times 10^{-2}$ \qquad \hfill|
\hfill \quad 0.10 \qquad \hfill|
\hfill \quad 0.13 \qquad \hfill\cr
\hfill \quad 1 \qquad \hfill|
\hfill \quad 1.48 \qquad \hfill|
\hfill \quad 0.15 \qquad \hfill|
\hfill \quad 0.58 \qquad \hfill|
\hfill \quad 36.3 \qquad \hfill|
\hfill \quad 6.0 \qquad \hfill|
\hfill \quad $1.0\times 10^{-3}$ \qquad \hfill|
\hfill \quad 0.093 \qquad \hfill|
\hfill \quad 0.12 \qquad \hfill\cr
\hfill \quad 2 \qquad \hfill|
\hfill \quad 1.59 \qquad \hfill|
\hfill \quad 0.082 \qquad \hfill|
\hfill \quad 0.30 \qquad \hfill|
\hfill \quad 30.4 \qquad \hfill|
\hfill \quad 1.5 \qquad \hfill|
\hfill \quad $1.1\times 10^{-6}$ \qquad \hfill|
\hfill \quad 0.054 \qquad \hfill|
\hfill \quad 0.077 \qquad \hfill\cr
\hfill \quad 3 \qquad \hfill|
\hfill \quad 1.78 \qquad \hfill|
\hfill \quad 0.046 \qquad \hfill|
\hfill \quad 0.15 \qquad \hfill|
\hfill \quad 15.1 \qquad \hfill|
\hfill \quad 0.11 \qquad \hfill|
\hfill \quad $9.4\times 10^{-12}$ \qquad \hfill|
\hfill \quad 0.029 \qquad \hfill|
\hfill \quad 0.035 \qquad \hfill\cr
\hfill \quad 5 \qquad \hfill|
\hfill \quad 1.90 \qquad \hfill|
\hfill \quad 0.020 \qquad \hfill|
\hfill \quad 0.020 \qquad \hfill|
\hfill \quad 3.56 \qquad \hfill|
\hfill \quad $3.7\times 10^{-4}$ \qquad \hfill|
\hfill \quad $9.8\times 10^{-24}$ \qquad \hfill|
\hfill \quad 0.0068 \qquad \hfill|
\hfill \quad 0.0086 \qquad \hfill
\endtable
\vskip -0.3cm
Here $L_x^*$ and $L$ are in units of $10^{44}$erg/s;
$n_0$,
$n(L>10^{43})$
and $n(L>10^{44})$
are in units of $10^{-6}h^3$Mpc$^{-3}$;
$j_{cl}$ and $j_{gas}$ are in
units of $10^{40}$erg/s/h$^{-3}$Mpc$^3$, and $j_{cl}$ may be larger
than $j_{gas}$ due to the inaccuracy of the Schechter fit.
\vfill\eject

\centerline {{\bf Table 1b.} Parameters of Schechter fits for the
 X-ray cluster luminosity function}
\vskip -0.3cm
\centerline {in $0.3-3.5$keV band}
\vskip -0.3cm
\centerline {X-ray Cluster Core Luminosity ($<0.5h^{-1}$Mpc)}
\begintable
\hfill \quad $z$ \qquad\hfill|
\hfill \quad $\alpha$ \qquad\hfill|
\hfill \quad $L_x^*$ \qquad\hfill|
\hfill \quad $n_0$ \qquad\hfill|
\hfill \quad $n(L>10^{43})$ \qquad\hfill|
\hfill \quad $n(L>10^{44})$ \qquad\hfill|
\hfill \quad $j_{cl}$ \qquad\hfill|
\hfill \quad $j_{gas}$ \qquad\hfill\cr
\hfill \quad 0 \qquad \hfill|
\hfill \quad 1.45 \qquad \hfill|
\hfill \quad 0.22 \qquad \hfill|
\hfill \quad 5.86 \qquad \hfill|
\hfill \quad 1.7 \qquad \hfill|
\hfill \quad $2.4\times 10^{-3}$ \qquad \hfill|
\hfill \quad 0.021 \qquad \hfill|
\hfill \quad 0.060 \qquad \hfill\cr
\hfill \quad 0.2 \qquad \hfill|
\hfill \quad 1.48 \qquad \hfill|
\hfill \quad 0.25 \qquad \hfill|
\hfill \quad 6.05 \qquad \hfill|
\hfill \quad 2.1 \qquad \hfill|
\hfill \quad $4.7\times 10^{-3}$ \qquad \hfill|
\hfill \quad 0.026 \qquad \hfill|
\hfill \quad 0.070 \qquad \hfill\cr
\hfill \quad 0.5 \qquad \hfill|
\hfill \quad 1.45 \qquad \hfill|
\hfill \quad 0.26 \qquad \hfill|
\hfill \quad 9.32 \qquad \hfill|
\hfill \quad 2.4 \qquad \hfill|
\hfill \quad $9.3\times 10^{-3}$ \qquad \hfill|
\hfill \quad 0.039 \qquad \hfill|
\hfill \quad 0.075 \qquad \hfill\cr
\hfill \quad 0.7 \qquad \hfill|
\hfill \quad 1.46 \qquad \hfill|
\hfill \quad 0.24 \qquad \hfill|
\hfill \quad 9.66 \qquad \hfill|
\hfill \quad 3.2 \qquad \hfill|
\hfill \quad $6.3\times 10^{-3}$ \qquad \hfill|
\hfill \quad 0.038 \qquad \hfill|
\hfill \quad 0.069 \qquad \hfill\cr
\hfill \quad 1 \qquad \hfill|
\hfill \quad 1.35 \qquad \hfill|
\hfill \quad 0.22 \qquad \hfill|
\hfill \quad 8.40 \qquad \hfill|
\hfill \quad 2.4 \qquad \hfill|
\hfill \quad $4.0\times 10^{-3}$ \qquad \hfill|
\hfill \quad 0.026 \qquad \hfill|
\hfill \quad 0.059 \qquad \hfill\cr
\hfill \quad 2 \qquad \hfill|
\hfill \quad 1.38 \qquad \hfill|
\hfill \quad 0.14 \qquad \hfill|
\hfill \quad 5.25 \qquad \hfill|
\hfill \quad 0.78 \qquad \hfill|
\hfill \quad $1.0\times 10^{-4}$ \qquad \hfill|
\hfill \quad 0.011 \qquad \hfill|
\hfill \quad 0.020 \qquad \hfill\cr
\hfill \quad 3 \qquad \hfill|
\hfill \quad 1.60 \qquad \hfill|
\hfill \quad 0.13 \qquad \hfill|
\hfill \quad 3.02 \qquad \hfill|
\hfill \quad 0.38 \qquad \hfill|
\hfill \quad $1.9\times 10^{-5}$ \qquad \hfill|
\hfill \quad 0.0087 \qquad \hfill|
\hfill \quad 0.0034 \qquad \hfill\cr
\hfill \quad 5 \qquad \hfill|
\hfill \quad 1.90 \qquad \hfill|
\hfill \quad 0.010 \qquad \hfill|
\hfill \quad 0.03 \qquad \hfill|
\hfill \quad $6.4\times 10^{-9}$ \qquad \hfill|
\hfill \quad $4.4\times 10^{-26}$ \qquad \hfill|
\hfill \quad $2.8\times 10^{-5}$ \qquad \hfill|
\hfill \quad $2.6\times 10^{-5}$ \qquad \hfill
\endtable
\centerline {X-ray Cluster Total Luminosity ($<1h^{-1}$Mpc)}
\begintable
\hfill \quad $z$ \qquad\hfill|
\hfill \quad $\alpha$ \qquad\hfill|
\hfill \quad $L_x^*$ \qquad\hfill|
\hfill \quad $n_0$ \qquad\hfill|
\hfill \quad $n(L>10^{43})$ \qquad\hfill|
\hfill \quad $n(L>10^{44})$ \qquad\hfill|
\hfill \quad $j_{cl}$ \qquad\hfill|
\hfill \quad $j_{gas}$ \qquad\hfill\cr
\hfill \quad 0 \qquad \hfill|
\hfill \quad 1.42 \qquad \hfill|
\hfill \quad 0.44 \qquad \hfill|
\hfill \quad 6.37 \qquad \hfill|
\hfill \quad 4.1 \qquad \hfill|
\hfill \quad $6.0\times 10^{-2}$ \qquad \hfill|
\hfill \quad 0.043 \qquad \hfill|
\hfill \quad 0.060 \qquad \hfill\cr
\hfill \quad 0.2 \qquad \hfill|
\hfill \quad 1.43 \qquad \hfill|
\hfill \quad 0.48 \qquad \hfill|
\hfill \quad 6.94 \qquad \hfill|
\hfill \quad 4.9 \qquad \hfill|
\hfill \quad $8.5\times 10^{-2}$ \qquad \hfill|
\hfill \quad 0.052 \qquad \hfill|
\hfill \quad 0.070 \qquad \hfill\cr
\hfill \quad 0.5 \qquad \hfill|
\hfill \quad 1.37 \qquad \hfill|
\hfill \quad 0.32 \qquad \hfill|
\hfill \quad 14.2 \qquad \hfill|
\hfill \quad 6.3 \qquad \hfill|
\hfill \quad $4.2\times 10^{-2}$ \qquad \hfill|
\hfill \quad 0.065 \qquad \hfill|
\hfill \quad 0.075 \qquad \hfill\cr
\hfill \quad 0.7 \qquad \hfill|
\hfill \quad 1.31 \qquad \hfill|
\hfill \quad 0.13 \qquad \hfill|
\hfill \quad 30.5 \qquad \hfill|
\hfill \quad 4.1 \qquad \hfill|
\hfill \quad $3.7\times 10^{-4}$ \qquad \hfill|
\hfill \quad 0.052 \qquad \hfill|
\hfill \quad 0.069 \qquad \hfill\cr
\hfill \quad 1 \qquad \hfill|
\hfill \quad 1.36 \qquad \hfill|
\hfill \quad 0.16 \qquad \hfill|
\hfill \quad 22.4 \qquad \hfill|
\hfill \quad 4.2 \qquad \hfill|
\hfill \quad $1.3\times 10^{-3}$ \qquad \hfill|
\hfill \quad 0.050 \qquad \hfill|
\hfill \quad 0.059 \qquad \hfill\cr
\hfill \quad 2 \qquad \hfill|
\hfill \quad 1.24 \qquad \hfill|
\hfill \quad 0.13 \qquad \hfill|
\hfill \quad 9.59 \qquad \hfill|
\hfill \quad 1.3 \qquad \hfill|
\hfill \quad $1.3\times 10^{-4}$ \qquad \hfill|
\hfill \quad 0.015 \qquad \hfill|
\hfill \quad 0.020 \qquad \hfill\cr
\hfill \quad 3 \qquad \hfill|
\hfill \quad 1.60 \qquad \hfill|
\hfill \quad 0.35 \qquad \hfill|
\hfill \quad 0.77 \qquad \hfill|
\hfill \quad $2.5\times 10^{-3}$ \qquad \hfill|
\hfill \quad $6.0\times 10^{-16}$ \qquad \hfill|
\hfill \quad 0.0060 \qquad \hfill|
\hfill \quad 0.0034 \qquad \hfill\cr
\hfill \quad 5 \qquad \hfill|
\hfill \quad 1.90 \qquad \hfill|
\hfill \quad 0.01 \qquad \hfill|
\hfill \quad 0.050 \qquad \hfill|
\hfill \quad $1.1\times 10^{-8}$ \qquad \hfill|
\hfill \quad $7.4\times 10^{-26}$ \qquad \hfill|
\hfill \quad $4.8\times 10^{-5}$ \qquad \hfill|
\hfill \quad $2.6\times 10^{-5}$ \qquad \hfill
\endtable
\vfill\eject

\centerline {{\bf Table 1c.} Parameters of Schechter fits for the
 X-ray cluster luminosity function}
\vskip -0.3cm
\centerline {in $0.5-4.5$keV band}
\vskip -0.3cm
\centerline {X-ray Cluster Core Luminosity ($<0.5h^{-1}$Mpc)}
\begintable
\hfill \quad $z$ \qquad\hfill|
\hfill \quad $\alpha$ \qquad\hfill|
\hfill \quad $L_x^*$ \qquad\hfill|
\hfill \quad $n_0$ \qquad\hfill|
\hfill \quad $n(L>10^{43})$ \qquad\hfill|
\hfill \quad $n(L>10^{44})$ \qquad\hfill|
\hfill \quad $j_{cl}$ \qquad\hfill|
\hfill \quad $j_{gas}$ \qquad\hfill\cr
\hfill \quad 0 \qquad \hfill|
\hfill \quad 1.39 \qquad \hfill|
\hfill \quad 0.21 \qquad \hfill|
\hfill \quad 6.32 \qquad \hfill|
\hfill \quad 1.7 \qquad \hfill|
\hfill \quad $2.2\times 10^{-3}$ \qquad \hfill|
\hfill \quad 0.019 \qquad \hfill|
\hfill \quad 0.053 \qquad \hfill\cr
\hfill \quad 0.2 \qquad \hfill|
\hfill \quad 1.44 \qquad \hfill|
\hfill \quad 0.26 \qquad \hfill|
\hfill \quad 5.74 \qquad \hfill|
\hfill \quad 2.1 \qquad \hfill|
\hfill \quad $5.8\times 10^{-3}$ \qquad \hfill|
\hfill \quad 0.024 \qquad \hfill|
\hfill \quad 0.059 \qquad \hfill\cr
\hfill \quad 0.5 \qquad \hfill|
\hfill \quad 1.46 \qquad \hfill|
\hfill \quad 0.57 \qquad \hfill|
\hfill \quad 4.65 \qquad \hfill|
\hfill \quad 4.0 \qquad \hfill|
\hfill \quad $9.4\times 10^{-2}$ \qquad \hfill|
\hfill \quad 0.044 \qquad \hfill|
\hfill \quad 0.061 \qquad \hfill\cr
\hfill \quad 0.7 \qquad \hfill|
\hfill \quad 1.39 \qquad \hfill|
\hfill \quad 0.16 \qquad \hfill|
\hfill \quad 12.1 \qquad \hfill|
\hfill \quad 2.2 \qquad \hfill|
\hfill \quad $6.7\times 10^{-4}$ \qquad \hfill|
\hfill \quad 0.028 \qquad \hfill|
\hfill \quad 0.053 \qquad \hfill\cr
\hfill \quad 1 \qquad \hfill|
\hfill \quad 1.34 \qquad \hfill|
\hfill \quad 0.13 \qquad \hfill|
\hfill \quad 10.9 \qquad \hfill|
\hfill \quad 1.4 \qquad \hfill|
\hfill \quad $1.2\times 10^{-4}$ \qquad \hfill|
\hfill \quad 0.019 \qquad \hfill|
\hfill \quad 0.042 \qquad \hfill\cr
\hfill \quad 2 \qquad \hfill|
\hfill \quad 1.33 \qquad \hfill|
\hfill \quad 0.054 \qquad \hfill|
\hfill \quad 9.30 \qquad \hfill|
\hfill \quad 0.18 \qquad \hfill|
\hfill \quad $7.2\times 10^{-10}$ \qquad \hfill|
\hfill \quad 0.0068 \qquad \hfill|
\hfill \quad 0.012 \qquad \hfill\cr
\hfill \quad 3 \qquad \hfill|
\hfill \quad 1.58 \qquad \hfill|
\hfill \quad 0.025 \qquad \hfill|
\hfill \quad 5.2 \qquad \hfill|
\hfill \quad $3.5\times 10^{-3}$ \qquad \hfill|
\hfill \quad $2.9\times 10^{-20}$ \qquad \hfill|
\hfill \quad $2.7\times 10^{-4}$ \qquad \hfill|
\hfill \quad $1.5\times 10^{-3}$ \qquad \hfill\cr
\hfill \quad 5 \qquad \hfill|
\hfill \quad 1.65 \qquad \hfill|
\hfill \quad 0.09 \qquad \hfill|
\hfill \quad 1.60 \qquad \hfill|
\hfill \quad $9.2\times 10^{-2}$ \qquad \hfill|
\hfill \quad $1.7\times 10^{-7}$ \qquad \hfill|
\hfill \quad 0.004 \qquad \hfill|
\hfill \quad $5.4\times 10^{-6}$ \qquad \hfill
\endtable
\centerline {X-ray Cluster Total Luminosity ($<1h^{-1}$Mpc)}
\begintable
\hfill \quad $z$ \qquad\hfill|
\hfill \quad $\alpha$ \qquad\hfill|
\hfill \quad $L_x^*$ \qquad\hfill|
\hfill \quad $n_0$ \qquad\hfill|
\hfill \quad $n(L>10^{43})$ \qquad\hfill|
\hfill \quad $n(L>10^{44})$ \qquad\hfill|
\hfill \quad $j_{cl}$ \qquad\hfill|
\hfill \quad $j_{gas}$ \qquad\hfill\cr
\hfill \quad 0 \qquad \hfill|
\hfill \quad 1.36 \qquad \hfill|
\hfill \quad 0.44 \qquad \hfill|
\hfill \quad 6.46 \qquad \hfill|
\hfill \quad 4.0 \qquad \hfill|
\hfill \quad $6.4\times 10^{-2}$ \qquad \hfill|
\hfill \quad 0.040 \qquad \hfill|
\hfill \quad 0.053 \qquad \hfill\cr
\hfill \quad 0.2 \qquad \hfill|
\hfill \quad 1.42 \qquad \hfill|
\hfill \quad 0.58 \qquad \hfill|
\hfill \quad 5.23 \qquad \hfill|
\hfill \quad 4.4 \qquad \hfill|
\hfill \quad 0.11 \qquad \hfill|
\hfill \quad 0.047 \qquad \hfill|
\hfill \quad 0.059 \qquad \hfill\cr
\hfill \quad 0.5 \qquad \hfill|
\hfill \quad 1.40 \qquad \hfill|
\hfill \quad 0.52 \qquad \hfill|
\hfill \quad 7.92 \qquad \hfill|
\hfill \quad 5.9 \qquad \hfill|
\hfill \quad 0.13 \qquad \hfill|
\hfill \quad 0.061 \qquad \hfill|
\hfill \quad 0.061 \qquad \hfill\cr
\hfill \quad 0.7 \qquad \hfill|
\hfill \quad 1.32 \qquad \hfill|
\hfill \quad 0.14 \qquad \hfill|
\hfill \quad 20.6 \qquad \hfill|
\hfill \quad 3.1 \qquad \hfill|
\hfill \quad $4.6\times 10^{-4}$ \qquad \hfill|
\hfill \quad 0.038 \qquad \hfill|
\hfill \quad 0.053 \qquad \hfill\cr
\hfill \quad 1 \qquad \hfill|
\hfill \quad 1.31 \qquad \hfill|
\hfill \quad 0.12 \qquad \hfill|
\hfill \quad 22.4 \qquad \hfill|
\hfill \quad 2.6 \qquad \hfill|
\hfill \quad $1.3\times 10^{-4}$ \qquad \hfill|
\hfill \quad 0.035 \qquad \hfill|
\hfill \quad 0.042 \qquad \hfill\cr
\hfill \quad 2 \qquad \hfill|
\hfill \quad 1.38 \qquad \hfill|
\hfill \quad 0.051 \qquad \hfill|
\hfill \quad 11.7 \qquad \hfill|
\hfill \quad 0.18 \qquad \hfill|
\hfill \quad $2.4\times 10^{-10}$ \qquad \hfill|
\hfill \quad 0.0086 \qquad \hfill|
\hfill \quad 0.012 \qquad \hfill\cr
\hfill \quad 3 \qquad \hfill|
\hfill \quad 1.58 \qquad \hfill|
\hfill \quad 0.035 \qquad \hfill|
\hfill \quad 5.2 \qquad \hfill|
\hfill \quad $1.7\times 10^{-2}$ \qquad \hfill|
\hfill \quad $4.3\times 10^{-15}$ \qquad \hfill|
\hfill \quad $3.8\times 10^{-3}$ \qquad \hfill|
\hfill \quad $1.5\times 10^{-3}$ \qquad \hfill\cr
\hfill \quad 5 \qquad \hfill|
\hfill \quad 1.43 \qquad \hfill|
\hfill \quad 5.95 \qquad \hfill|
\hfill \quad 0.47 \qquad \hfill|
\hfill \quad $3.1\times 10^{-2}$ \qquad \hfill|
\hfill \quad $8.8\times 10^{-8}$ \qquad \hfill|
\hfill \quad 0.044 \qquad \hfill|
\hfill \quad $5.4\times 10^{-6}$ \qquad \hfill
\endtable
\vfill\eject

\centerline {{\bf Table 1d.} Parameters of Schechter fits for the
 X-ray cluster luminosity function}
\vskip -0.3cm
\centerline {in $2-10$keV band}
\vskip -0.3cm
\centerline {X-ray Cluster Core Luminosity ($<0.5h^{-1}$Mpc)}
\begintable
\hfill \quad $z$ \qquad\hfill|
\hfill \quad $\alpha$ \qquad\hfill|
\hfill \quad $L_x^*$ \qquad\hfill|
\hfill \quad $n_0$ \qquad\hfill|
\hfill \quad $n(L>10^{43})$ \qquad\hfill|
\hfill \quad $n(L>10^{44})$ \qquad\hfill|
\hfill \quad $j_{cl}$ \qquad\hfill|
\hfill \quad $j_{gas}$ \qquad\hfill\cr
\hfill \quad 0 \qquad \hfill|
\hfill \quad 1.22 \qquad \hfill|
\hfill \quad 0.088 \qquad \hfill|
\hfill \quad 9.84 \qquad \hfill|
\hfill \quad 0.67 \qquad \hfill|
\hfill \quad $2.4\times 10^{-6}$ \qquad \hfill|
\hfill \quad 0.011 \qquad \hfill|
\hfill \quad 0.026 \qquad \hfill\cr
\hfill \quad 0.2 \qquad \hfill|
\hfill \quad 1.37 \qquad \hfill|
\hfill \quad 0.16 \qquad \hfill|
\hfill \quad 5.01 \qquad \hfill|
\hfill \quad 0.92 \qquad \hfill|
\hfill \quad $2.9\times 10^{-4}$ \qquad \hfill|
\hfill \quad 0.011 \qquad \hfill|
\hfill \quad 0.024 \qquad \hfill\cr
\hfill \quad 0.5 \qquad \hfill|
\hfill \quad 1.34 \qquad \hfill|
\hfill \quad 0.15 \qquad \hfill|
\hfill \quad 6.70 \qquad \hfill|
\hfill \quad 1.1 \qquad \hfill|
\hfill \quad $2.5\times 10^{-4}$ \qquad \hfill|
\hfill \quad 0.014 \qquad \hfill|
\hfill \quad 0.019 \qquad \hfill\cr
\hfill \quad 0.7 \qquad \hfill|
\hfill \quad 1.35 \qquad \hfill|
\hfill \quad 0.17 \qquad \hfill|
\hfill \quad 3.13 \qquad \hfill|
\hfill \quad 0.63 \qquad \hfill|
\hfill \quad $2.9\times 10^{-4}$ \qquad \hfill|
\hfill \quad 0.0074 \qquad \hfill|
\hfill \quad 0.013 \qquad \hfill\cr
\hfill \quad 1 \qquad \hfill|
\hfill \quad 1.20 \qquad \hfill|
\hfill \quad 0.10 \qquad \hfill|
\hfill \quad 3.88 \qquad \hfill|
\hfill \quad 0.34 \qquad \hfill|
\hfill \quad $4.4\times 10^{-6}$ \qquad \hfill|
\hfill \quad 0.0045 \qquad \hfill|
\hfill \quad 0.0078 \qquad \hfill\cr
\hfill \quad 2 \qquad \hfill|
\hfill \quad 1.17 \qquad \hfill|
\hfill \quad 0.081 \qquad \hfill|
\hfill \quad 0.79 \qquad \hfill|
\hfill \quad $4.7\times 10^{-2}$ \qquad \hfill|
\hfill \quad $7.4\times 10^{-8}$ \qquad \hfill|
\hfill \quad $7.0\times 10^{-4}$ \qquad \hfill|
\hfill \quad $7.8\times 10^{-4}$ \qquad \hfill\cr
\hfill \quad 3 \qquad \hfill|
\hfill \quad 1.13  \qquad \hfill|
\hfill \quad 0.018 \qquad \hfill|
\hfill \quad 0.14 \qquad \hfill|
\hfill \quad $2.9\times 10^{-5}$ \qquad \hfill|
\hfill \quad $5.1\times 10^{-23}$ \qquad \hfill|
\hfill \quad $2.8\times 10^{-5}$ \qquad \hfill|
\hfill \quad $3.1\times 10^{-5}$ \qquad \hfill\cr
\hfill \quad 5 \qquad \hfill|
\hfill \quad 1.80 \qquad \hfill|
\hfill \quad 0.004\qquad \hfill|
\hfill \quad 0.01 \qquad \hfill|
\hfill \quad $1.8\times 10^{-16}$ \qquad \hfill|
\hfill \quad $1.3\times 10^{-26}$ \qquad \hfill|
\hfill \quad $1.8\times 10^{-6}$ \qquad \hfill|
\hfill \quad $2.5\times 10^{-10}$ \qquad \hfill
\endtable
\centerline {X-ray Cluster Total Luminosity ($<1h^{-1}$Mpc)}
\begintable
\hfill \quad $z$ \qquad\hfill|
\hfill \quad $\alpha$ \qquad\hfill|
\hfill \quad $L_x^*$ \qquad\hfill|
\hfill \quad $n_0$ \qquad\hfill|
\hfill \quad $n(L>10^{43})$ \qquad\hfill|
\hfill \quad $n(L>10^{44})$ \qquad\hfill|
\hfill \quad $j_{cl}$ \qquad\hfill|
\hfill \quad $j_{gas}$ \qquad\hfill\cr
\hfill \quad 0 \qquad \hfill|
\hfill \quad 1.19 \qquad \hfill|
\hfill \quad 0.18 \qquad \hfill|
\hfill \quad 9.40 \qquad \hfill|
\hfill \quad 2.0 \qquad \hfill|
\hfill \quad $1.7\times 10^{-3}$ \qquad \hfill|
\hfill \quad 0.020 \qquad \hfill|
\hfill \quad 0.026 \qquad \hfill\cr
\hfill \quad 0.2 \qquad \hfill|
\hfill \quad 1.35 \qquad \hfill|
\hfill \quad 0.30 \qquad \hfill|
\hfill \quad 5.88 \qquad \hfill|
\hfill \quad 2.4 \qquad \hfill|
\hfill \quad $1.3\times 10^{-2}$ \qquad \hfill|
\hfill \quad 0.024 \qquad \hfill|
\hfill \quad 0.024 \qquad \hfill\cr
\hfill \quad 0.5 \qquad \hfill|
\hfill \quad 1.31 \qquad \hfill|
\hfill \quad 0.25 \qquad \hfill|
\hfill \quad 5.01 \qquad \hfill|
\hfill \quad 1.7 \qquad \hfill|
\hfill \quad $5.1\times 10^{-3}$ \qquad \hfill|
\hfill \quad 0.016 \qquad \hfill|
\hfill \quad 0.019 \qquad \hfill\cr
\hfill \quad 0.7 \qquad \hfill|
\hfill \quad 1.32 \qquad \hfill|
\hfill \quad 0.20 \qquad \hfill|
\hfill \quad 2.82 \qquad \hfill|
\hfill \quad 0.71 \qquad \hfill|
\hfill \quad $8.1\times 10^{-4}$ \qquad \hfill|
\hfill \quad 0.010 \qquad \hfill|
\hfill \quad 0.013 \qquad \hfill\cr
\hfill \quad 1 \qquad \hfill|
\hfill \quad 1.13 \qquad \hfill|
\hfill \quad 0.17 \qquad \hfill|
\hfill \quad 3.14 \qquad \hfill|
\hfill \quad 0.63 \qquad \hfill|
\hfill \quad $4.4\times 10^{-4}$ \qquad \hfill|
\hfill \quad 0.0058 \qquad \hfill|
\hfill \quad 0.0078 \qquad \hfill\cr
\hfill \quad 2 \qquad \hfill|
\hfill \quad 1.10 \qquad \hfill|
\hfill \quad 0.10 \qquad \hfill|
\hfill \quad 1.25 \qquad \hfill|
\hfill \quad 0.11 \qquad \hfill|
\hfill \quad $1.8\times 10^{-6}$ \qquad \hfill|
\hfill \quad $1.3\times 10^{-3}$ \qquad \hfill|
\hfill \quad $7.8\times 10^{-4}$ \qquad \hfill\cr
\hfill \quad 3 \qquad \hfill|
\hfill \quad 1.03  \qquad \hfill|
\hfill \quad 0.036 \qquad \hfill|
\hfill \quad 0.14 \qquad \hfill|
\hfill \quad $1.0\times 10^{-3}$ \qquad \hfill|
\hfill \quad $1.7\times 10^{-15}$ \qquad \hfill|
\hfill \quad $5.2\times 10^{-5}$ \qquad \hfill|
\hfill \quad $3.1\times 10^{-5}$ \qquad \hfill\cr
\hfill \quad 5 \qquad \hfill|
\hfill \quad 1.80 \qquad \hfill|
\hfill \quad 0.005\qquad \hfill|
\hfill \quad 0.01 \qquad \hfill|
\hfill \quad $3.8\times 10^{-14}$ \qquad \hfill|
\hfill \quad $1.5\times 10^{-26}$ \qquad \hfill|
\hfill \quad $2.3\times 10^{-6}$ \qquad \hfill|
\hfill \quad $2.5\times 10^{-10}$ \qquad \hfill
\endtable

\vfill\eject\end